# Particles Flux of Ultrahigh Energy from a Galactic Plane

A.A. MIKHAILOV

Yu.G. Shafer Institute of Cosmophysical Research and Aeronomy of SB RAS, 677980, Yakutsk, Russia
mikhailov@ikfia.ysn.ru

## ABSTRACT

Arrival directions of particles were analyzed according to the arrays of extensive air showers (EAS) data. A new method of sources search and anisotropy of arrival directions particles is suggested. There was found the particles flux of ultrahigh energy from the galactic plane that exceeds the expected number of particles in the case of isotropy more than 4σ and 5σ.

## 1. INTRODUCTION

The origin of cosmic rays of ultrahigh energy is not found out yet. It can be generated at explosions of supernovas (e.g., Berezhko et al. 2003), in pulsars (e.g. Giler & Lipski 2001; Olinto et al., 1999; Hong-Bo et al.,2009) in the nucleus of active galaxies (The P. Auger Collaboration, 2008; Berezinsky et al, 2006) and etc.

## 2. METHOD

We suggest a new method of sources search. For each proposed source of particles we shall define number of events inside of angular distance R from it. We selected as sources of particles where the number of particles is above a certain threshold (≥n) inside <R from it. If the distance between two nearest sources is less <2R we accept crossed regions as one region. We determine observed number of events inside of this region. Then we estimate chance probability of observed number of events in the given region from the common number of the registered particles. If the observed number of events around of proposed sources is not chance these sources we consider as their probable sources.

## 3. RESULTS

In papers (Mikhailov, 1999; Becker et al., 2004) and etc. we found a correlation between arrival directions of ultrahigh particles energy according to Yakutsk EAS array data and pulsars located along lines of a large-scale regular magnetic field of the Orion arm. It was observed the correlation between them at angular distance R<6°. In this paper we searched for correlation between arrival directions of ultrahigh energy particles and pulsars in all celestial sphere inside at the above-mentioned angular distance R.

We analyzed an arrival directions of ultrahigh energy particles according to Yakutsk EAS array data at energy $E>8\times10^{18}$ eV, to P. Auger EAS array data at $E>5.7\times10^{19}$ eV (The P. Auger Collaboration, 2010). Data of other arrays at $E>4\times10^{19}$ eV is not considered because of their insufficient statistics.

At first we consider an arrival directions of 898 particles according to Yakutsk EAS array data at energy $E>8\times10^{18}$ eV. The pulsars from the catalogue (Manchester et al., 2005) are shown which around at $R<6°$ the number of particles more $n\geq10$ on the map of equal exposition of celestial sphere where the equal number of particles is expected from equal area (Fig.1). We have received 3 groups of pulsars after connection of crossed regions with angular distance $<2R$ between their. There were found 37 in 1-st group, 4 in 2-nd group, 6 pulsars in 3-rd group. There are 178, 27 and 14 particles in these groups of pulsars accordingly. The chance probability to find out such number of events in these groups of pulsars from 898 particles is equal $P_1 \sim 3.10^{-6}$, $P_2 \sim 0.8$ and $P_3 \sim 2.2\times10^{-3}$ according to (Efimov, Mikhailov, 1994).

The location of 1 and 3 groups of pulsars from which the statistically-significant number of particles is observed, are in a galactic plane. It is observed the maximal number of events $20\leq n\leq23$ around of pulsars 1-st groups J0141+6009, 0146+6145, 0147+5922, 0157+6212, 0215+6218. The given pulsars as probable sources of particles have been considered earlier (Mikhailov, 1999; Erlykin et al., 2002; Malov et al., 2010).

We note that there were found particles at $E\sim10^{18}$ eV near the center of the Galaxy according to array data AGASA (Hayashida et al., 1999) and Sydney (Bellido et al., 2001) where fluxes of particles are observed (Fig.1, the group 3).

We have considered arrival directions of particles according to array P. Auger at energy $E>5.7\times10^{19}$ eV. We have selected those pulsars around which $\geq3$ particles are observed inside of angular distance $R<6°$ - 24 pulsars match the given term (Fig.2, large black circles). 23 pulsars are inside at $R<6°$ from each other, except for pulsar J1332-3032. The chance probability to find 9 events around of the given groups of pulsars from 69 particles is equal $P\sim10^{-4}$. These pulsars also are near to a galactic plane (Fig.2).

Then we selected those pulsars around which $\geq2$ particles are observed at $R<6°$ (Fig.3, small black circles). Thus we found out 48 pulsars between a right ascension $180°<RA<240°$. The chance probability to find out 14 events in the given group of pulsars from 69 particles is equal $P \sim 10^{-6}$. These pulsars are near to a galactic plane at Output of the Orion arm (Fig.2).

We have estimated anisotropy of arrival directions of particles at considered energy. We draw line of a coordinate grid on declination $\delta$ and a right ascension RA through $5°$ on all celestial sphere, since the declination $\delta=0°$ and the right ascension $RA=0°$ have counted up the number of observed particles inside an angular distance of R around points where the above-stated coordinates are crossed.

We have chosen areas $R<6°$ around given points where the observed number of particles n was $n\geq16$ events for Yakutsk array. We have found 113 particles after the connection of crossed regions into one (Fig.1). The chance of probability to find out such number of events from general number 898 of particles is $P_1\sim2.10^{-5}$. The region of arrival directions particles with $n\geq16$ at region $45°<\delta<80°$ и $60°<RA<95°$ larger than region occupied by pulsars of 1-st group. It is caused by small number of pulsars in the given region – there are only 4 pulsars according to (Manchester et al., 2005), 3 pulsars from them are shown in Fig.1, around of 4-th pulsar the number of particles is small.

The maximum of particle distribution is at coordinates $\delta \sim 60°$, $RA\sim 0°\div 40°$ ($b \sim 0°$, $l \sim 117°\div136°$, where l - a galactic longitude). The number of particles around of considered coordinates at $R<6°$ is equal $n\geq18$. The peak of the given maximum of distribution is at $\delta=60°$ and $RA=25°$. It is observed around this point at $R<6°$ 23 particles (an expected number of events in case of isotropy is $n\sim9.5$). Chance probability to observe such number of events from 898 particles is $P\sim10^{-4}$. There is pulsar J0205+6449 with the short period of rotation $P_0=0.06$ sec on distance 3.2 kpc on angular distance $5.6°$ near this maximum which is the most probable source of particles according to (e.g. Giler & Lipski 2001; Olinto et al., 1999).

We considered regions around the given points with radius R<6° in case of P. Auger array where the number of observed number of particles was n≥2 events (Fig.2). We have found 16 particles after the connection of crossed regions (Fig.2, red crossed curve). The chance of probability to find out such number of events from 69 particles is equal $P_1 < 10^{-6}$. The maximum of this distribution is around at coordinates δ ~ -55° and RA ~ 205° ÷ 210° and δ ~ -35° ÷ -60°, RA ~ 205° (b ~0 ÷ 15°, l ~310°). Peak of the given maximum of distribution is at δ ~ -55° and RA=210° (b ~6°, l ~313°). There are observed 5 particles at R <6° around this point (an expected number of events in case of isotropy is n ~ 0.38). The chance of probability to observe such number of events from 69 particles is equal $P \sim 4.10^{-5}$. From the given maximum of distribution of particles on distance 3.1° and 5.6° there are pulsars J0205+6449, J1439-5501 with the short periods of rotation $P_0$ ~ 0.034 and 0.028 sec on distances 1.7 and 0.7 kpc from the Earth.

## 5. SUMMARY

According to arrays EAS Yakutsk and P.Auger fluxes of particles ultrahigh energy from a galactic plane and the near center of the Galaxy are found out and we come to conclusion that their probable sources are pulsars.

## REFERENCES

Becker P.A., Bisnovatyi-Kogan G.S., Casadei D., …, Mikhailov A.A., et al. Frontiers Cosmic Ray Research, NY, 2004, 161.
Bellido J.A., Clay R.W., Dawson B.R., Johston-Hollit M.. Astrop. Phys. 2001, 15, 167.
Berezhko E.G., Ksenofontov L.T., Ptuskin V.S. Astron. and Astroph. 2003, 410, 189.
Berezinsky V.S., Gazizov A., Grigorieva S.. Phys. Rev. D., 2006, 74, 043005.
Efimov N.N., Mikhailov A.A. Astropart. Phys., 1994, 2, 329.
Erlykin A.D., Mikhailov A.A., Wolfendale A.W. J. Phys.G.: Nucl. Part. Phys. 2002, 28, 2225.
Giler M., Lipsk M., Proc. 27-th ICRC, Hamburg, 2001, 6, 2092.
Hayashida N., Nagano M., Nishikawa D. et al.. Astropart. Phys. 1999, 10, 303.
Hong-Bo Hu, Hu H., Yuan Q., Wang B. et al., Astroph. J., 2009, 700, L170.
Malov I.E., Mikhailov A.A., Avramenko A.E. et al. Pulsars: Theory, Categories and Applications, NY, 2010, 198.
Manchester R.N., Hobbs G. B., Teoh A., & Hobbs M., Astroph. J., 2005, 129, 1993.
Mikhailov A.A.. Izv. AS, ser. Phys.,. 1999, 63, 556.
Mikhailov A.A. Proc.26-th ICRC, Salt Lake City, 1999, 3, 268.
Olinto A.V., Epstein R.I., Blasi P. Proc. 26-th ICRC, Salt Lake City, 1999, 4, 361.
The P. Auger Collaboration. Astropart. Phys., 2008, 29, 188.
The P. Auger Collaboration. Astropart. Phys., 2010, 34, 314.

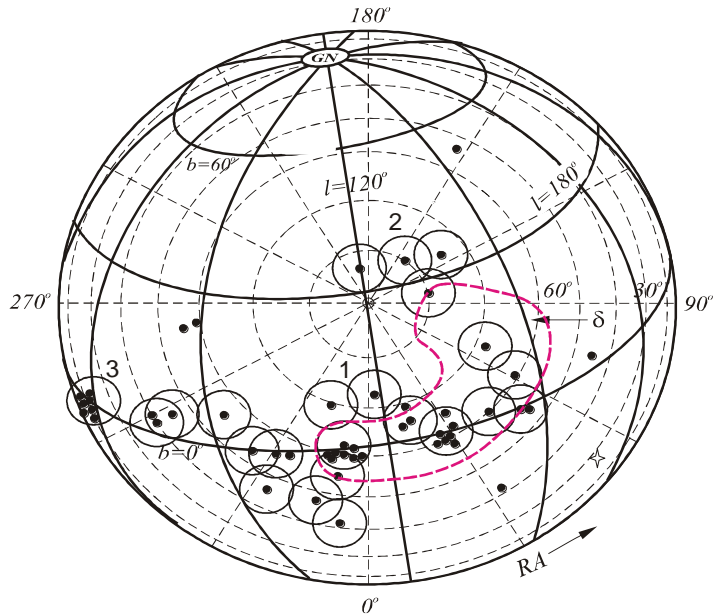

Fig.1. Distribution of pulsars to celestial sphere, Yakutsk: ● - pulsars around of which ≥10 particles were observed at R <6°, also it is shown in a circle around of pulsars with radius R=6°, δ- declination, RA - a right ascension, b - galactic latitude, l - a longitude. The red closed curve – a region around of points (5°×5°) where number of observed particles is n≥16 events.

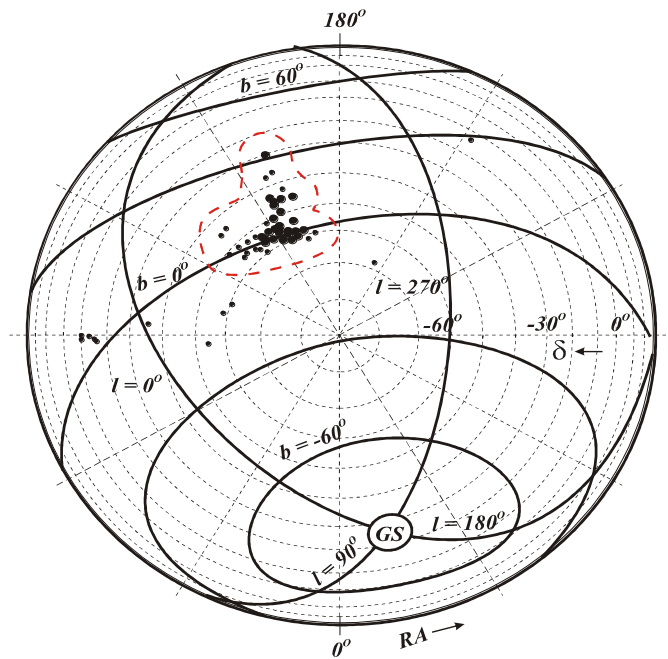

Fig.2. P. Auger: ●, ● - pulsars around of which ≥3 and ≥2 number of particles were observed at R <6°, other mark in same as Fig.1. The red closed curve – a region around of points (5°×5°) where number of observed particles is n≥2 events.